\documentclass[12pt, letterpaper]{article}
\usepackage[top=1in, bottom=1.5in, left=1in, right=1in]{geometry}
\usepackage[utf8]{inputenc}
\usepackage{booktabs}
\usepackage{hyperref}
\usepackage{graphicx}
\usepackage{gensymb}
\usepackage{natbib}
\usepackage{lineno}

\usepackage[usenames]{color}

\newcommand{\phil}[1]{\textcolor{black}{#1}}
\newcommand{\eric}[1]{\textcolor{black}{#1}}
\newcommand{\humna}[1]{\textcolor{black}{#1}}

\newcommand{\ttt}[1]{\texttt{#1}}
\def\ref@jnl#1{{\jnl@style#1}}%
\newcommand\aj{\ref@jnl{AJ}}
\newcommand\araa{\ref@jnl{ARA\&A}}
\newcommand\apj{\ref@jnl{ApJ}}
\newcommand\apjl{\ref@jnl{ApJL}}     
\newcommand\apjs{\ref@jnl{ApJS}}
\newcommand\ao{\ref@jnl{ApOpt}}
\newcommand\apss{\ref@jnl{Ap\&SS}}
\newcommand\aap{\ref@jnl{A\&A}}
\newcommand\aapr{\ref@jnl{A\&A~Rv}}
\newcommand\aaps{\ref@jnl{A\&AS}}
\newcommand\azh{\ref@jnl{AZh}}
\newcommand\baas{\ref@jnl{BAAS}}
\newcommand\icarus{\ref@jnl{Icarus}}
\newcommand\jrasc{\ref@jnl{JRASC}}
\newcommand\memras{\ref@jnl{MmRAS}}
\newcommand\mnras{\ref@jnl{MNRAS}}
\newcommand\pra{\ref@jnl{PhRvA}}
\newcommand\prb{\ref@jnl{PhRvB}}
\newcommand\prc{\ref@jnl{PhRvC}}
\newcommand\prd{\ref@jnl{PhRvD}}
\newcommand\pre{\ref@jnl{PhRvE}}
\newcommand\prl{\ref@jnl{PhRvL}}
\newcommand\pasp{\ref@jnl{PASP}}
\newcommand\pasj{\ref@jnl{PASJ}}
\newcommand\qjras{\ref@jnl{QJRAS}}
\newcommand\skytel{\ref@jnl{S\&T}}
\newcommand\solphys{\ref@jnl{SoPh}}
\newcommand\sovast{\ref@jnl{Soviet~Ast.}}
\newcommand\ssr{\ref@jnl{SSRv}}
\newcommand\zap{\ref@jnl{ZA}}
\newcommand\nat{\ref@jnl{Nature}}
\newcommand\iaucirc{\ref@jnl{IAUC}}
\newcommand\aplett{\ref@jnl{Astrophys.~Lett.}}
\newcommand\apspr{\ref@jnl{Astrophys.~Space~Phys.~Res.}}
\newcommand\bain{\ref@jnl{BAN}}
\newcommand\fcp{\ref@jnl{FCPh}}
\newcommand\gca{\ref@jnl{GeoCoA}}
\newcommand\grl{\ref@jnl{Geophys.~Res.~Lett.}}
\newcommand\jcp{\ref@jnl{JChPh}}
\newcommand\jgr{\ref@jnl{J.~Geophys.~Res.}}
\newcommand\jqsrt{\ref@jnl{JQSRT}}
\newcommand\memsai{\ref@jnl{MmSAI}}
\newcommand\nphysa{\ref@jnl{NuPhA}}
\newcommand\physrep{\ref@jnl{PhR}}
\newcommand\physscr{\ref@jnl{PhyS}}
\newcommand\planss{\ref@jnl{Planet.~Space~Sci.}}
\newcommand\procspie{\ref@jnl{Proc.~SPIE}}

\newcommand\actaa{\ref@jnl{AcA}}
\newcommand\caa{\ref@jnl{ChA\&A}}
\newcommand\cjaa{\ref@jnl{ChJA\&A}}
\newcommand\jcap{\ref@jnl{JCAP}}
\newcommand\na{\ref@jnl{NewA}}
\newcommand\nar{\ref@jnl{NewAR}}
\newcommand\pasa{\ref@jnl{PASA}}
\newcommand\rmxaa{\ref@jnl{RMxAA}}

\newcommand\maps{\ref@jnl{M\&PS}}
\newcommand\aas{\ref@jnl{AAS Meeting Abstracts}}
\newcommand\dps{\ref@jnl{AAS/DPS Meeting Abstracts}}

\title{A Big Sky Approach to Cadence Diplomacy}
\author{%
Knut Olsen (SMWLV), Marcella Di Criscienzo (SMWLV, TVS), \\ R. Lynne Jones (Solar System), Megan E. Schwamb (Solar System), \\ Hsing Wen ``Edward" Lin (Solar System),  Humna Awan (DESC), \\ Phil Marshall (DESC, Strong Lensing), Eric Gawiser (Galaxies, DESC), \\ Adam Bolton, Daniel Eisenstein}
\date{November 30, 2018}

\begin{document}

\maketitle

\begin{abstract}
The LSST survey was designed to deliver transformative scientific results for four primary objectives: constraining dark energy and dark matter, taking an inventory of the Solar System, exploring the transient optical sky, and mapping the Milky Way. While the LSST Wide-Fast-Deep survey and accompanying Deep Drilling and mini-surveys will be ground-breaking for each of these areas, there remain competing demands on the survey area, depth, and temporal coverage amid a desire to maximize all three.
In this white paper, we seek to address a principal source of tension between the different LSST science collaborations, that of the survey area and depth that they each need in the parts of the sky that they care about. We present simple tools which can be used to explore trades between the area surveyed by LSST and the number of visits available per field \phil{and then} use these tools to propose a change to the baseline survey strategy. Specifically, we propose to reconfigure the WFD footprint to consist of low-extinction regions (limited by galactic latitude), with the number of visits per field in WFD limited by the LSST Science Requirements Document (SRD) `design' goal, and suggest assignment of the remaining LSST visits to the full visible LSST sky. This proposal addresses concerns with the WFD footprint raised by the DESC (as 
$\sim25\%$ 
\eric{of the current baseline WFD region is not usable for dark energy}
science due to MW dust extinction), eases the time required for
the NES and SCP mini-surveys (since in our proposal they would partially fall into the modified WFD footprint), raises the number of visits previously assigned to the GP region, and 
\eric{increases the overlap with DESI and other Northern hemisphere follow-up facilities.}  
This proposal alleviates many of the current concerns of 
Science Collaborations that represent the four \eric{scientific} 
pillars of LSST and provides a Big Sky approach to cadence diplomacy.

\end{abstract}

\clearpage
\section{White Paper Information}
\begin{enumerate}
    \item {\bf Contact information:} Knut Olsen, kolsen@noao.edu; 
    \item {\bf Science Category:} All.
    \item {\bf Survey Type Category:} Wide-Fast-Deep, Mini survey
    \item {\bf Observing Strategy Category:} an integrated program with science that hinges on the combination of pointing and detailed 
	observing strategy
\end{enumerate}


\clearpage

\section{Scientific Motivation\label{sec:motivation}}


The current baseline LSST survey (e.g. \ttt{baseline2018a}) defines the main WFD area as the 18,000 deg$^2$ between $-62\degree<{\rm Dec}<+2\degree$, but avoiding the inner Galactic plane.  The baseline plan also includes $\sim$8,000 deg$^2$ designated as candidate "mini-survey areas with significantly reduced coverage, typically $\sim170$ (SCP/GP) to $\sim250$ (NES) visits per field in current simulations.
There are several clear opportunities for further scientific optimization of this plan, \humna{addressing concerns} such as:
\\
$\bullet$ The WFD area encompasses the Galactic Plane in the direction of the anticenter, where extinction by dust is higher than desired for extragalactic science cases.\\
$\bullet$ The region with $\rm{Dec}<-62\degree$, where the Magellanic Clouds and their satellite system reside, and where we expect significant structure in the Galactic halo, has no guaranteed coverage.\\
$\bullet$ Only 50\% of the Ecliptic Plane, where the large majority of Solar System objects reside, is covered by WFD, with the northern half designated a candidate mini-survey.\\
$\bullet$ While the WFD area includes the outer Galactic Plane, the inner Plane, which contains the vast majority of Galactic variable sources and is of high importance to the SMWLV Collaboration, would receive a limited number of visits.

These concerns can be summarized as \phil{that the baseline cadence has}  \textbf{1) a WFD footprint that needs to be better catered to the areas of highest interest, and 2) areas designated as candidate mini-surveys that receive insufficient visits for the target science.}
\phil{This summary points us to a starting point for a solution: {\it get more visits}. A simple way to achieve this is by dropping the requirement that we take two ``snap'' exposures per visit, and instead take just one exposure of around 30 seconds duration.}

\phil{Then, we note that} a fundamental issue is that the survey must make tradeoffs between area, depth, and temporal coverage, while the scientific interests of the LSST community desire to maximize all three.  For simplicity, we explored trades \phil{between} the area of the LSST survey vs.\ the number of 30\humna{s} 
visits per field (and hence its depth), leaving aside the question of the temporal cadence. We have developed a simple \ttt{Python} tool,
\ttt{LynneSim}\phil{, to help evaluate the potential for these trades.
This tool, available for all to use in  \href{https://github.com/LSSTScienceCollaborations/survey_strategy_wp}{the LSST Science Collaborations' fork of the survey strategy white paper repository},  is illustrated in example \ttt{Jupyter} notebooks  \href{http://ls.st/acl}{here} and \href{http://ls.st/tx3}{here}, 
and is \href{http://ls.st/o9s}{scriptable}.}

In order to test different cases, we estimate the total number of visits available. \textit{Assuming that the survey strategy drops snaps in favor of single \eric{exposures per visit}}, we have a total of $\sim$ 2.6M visits available over the lifetime of LSST. We use 90\% of these visits for our analysis, leaving 10\% for other mini-surveys that we do not consider here.

As a simple starting point, and a way to demonstrate our tool, we consider a survey that covers the entire sky visible from Cerro Pach\'{o}n, $-90\degree<\rm{Dec}<+32\degree$, $\sim3\pi$ steradians, which we call \humna{the} \textit{Big Sky}.  There is strong motivation for covering an area of this size: 1) it would maximize the use of cosmological probes for Dark Energy (see \humna{Lochner at al.,} the LSST DESC WFD \phil{cadence white paper}); 2) structures in the Milky Way halo and disk are large on the sky and highly non-uniform, hence the discovery potential increases proportional to survey area; 3) the area would include the full Ecliptic Plane and would maximize the ability to find new planets/minor planets in our Solar System; 4) we would \phil{maximize} overlap with northern photometric and spectroscopic surveys (see Bolton et al., ``Maximizing the Joint Science Return of LSST and DESI"), which would 
\eric{improve}
the \phil{calibration} of photometric redshifts and assign
velocities to halo structures found in the region; and 5) we would increase the ability to follow up transients and variables with \humna{spectroscopic facilities} 
in the northern hemisphere, where most of these resources are located.

Given full-sky coverage, the simplest approach 
\eric{is to distribute} 
LSST visits uniformly.
We find that over a ten-year survey, assigning equal coverage to the full visible sky far exceeds the \href{https://docushare.lsstcorp.org/docushare/dsweb/Get/LPM-17}{LSST SRD} stretch goal of 20,000 $\deg^2$ for WFD, but reaches only 2/3 of the design goal of 825 median visits per field in the WFD, not enough to put the ``D'' in WFD. 
\eric{Hence we must} 
distribute the visits non-uniformly in order to satisfy the LSST \phil{science} requirements.

\begin{figure}[!h]
\hspace*{1.8em}
\begin{minipage}{0.45\paperwidth}
	\includegraphics[trim={105 105 105 105}, clip=false, width=0.27\paperwidth]
	{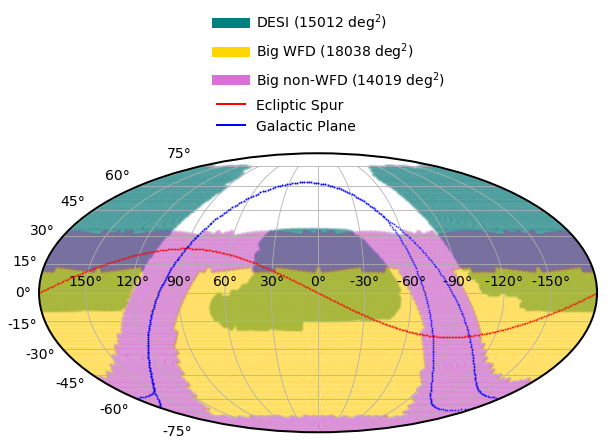}
\end{minipage}\
\hspace*{-2em}
\begin{minipage}{0.45\paperwidth}
	\includegraphics[trim={105 105 105 105}, clip=false, width=0.27\paperwidth]
	{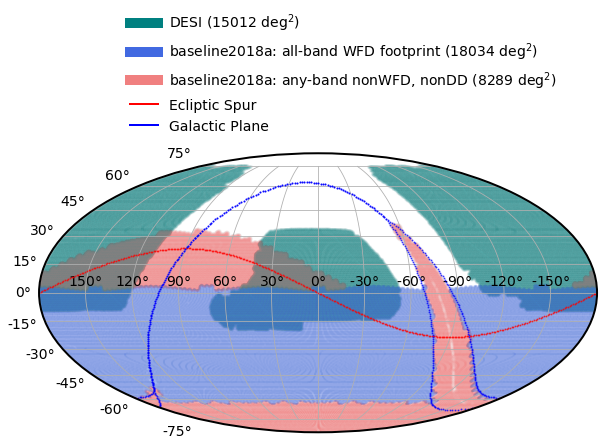}
\end{minipage}\
\vspace*{4em}
\caption{\humna{
    \textit{Left:} Our proposed Big Sky footprint: yellow fields denote our \phil{recommended} expanded WFD footprint while the purple fields represent the mini-surveys in the extended footprint.
    \textit{Right:} Footprint from \ttt{baseline2018a} for WFD (blue) and all the mini-surveys aside from the DDFs (coral red). Both plots show overlap the DESI footprint (aqua green), demonstrating that our Big Sky footprint significantly increases the overlap with DESI (5912 deg$^2$ for WFD and 4538 deg$^2$ for non-WFD) vs. \ttt{baseline2018a} (3739 deg$^2$ for WFD and 2233 deg$^2$ for non-WFD).
        }
    }
    \label{fig:coverage}
\end{figure}

\textbf{Our proposed solution is to \phil{instead} define WFD by avoiding the Galactic plane ($|b|>15\degree$) while extending symmetrically north and south far enough to meet the LSST SRD requirement of 18,000 $\deg^2$, which results in a WFD footprint that covers $-72.25\degree<\rm{Dec}<+12.4\degree$ (see \autoref{fig:coverage}).  We assign 825 visits per field to WFD, meeting the LSST SRD requirement of 825 median visits per field, and distribute the remaining visits to the non-WFD Big Sky fields.  
These non-WFD fields then receive $\sim250$ visits each, or 30\% of the WFD visits, which is larger than the current baseline strategy for most parts of the sky.}  

\begin{figure}[!h]
    \centering
    \includegraphics[width=0.6\columnwidth, trim={0cm 0cm 0cm 0cm}, clip]{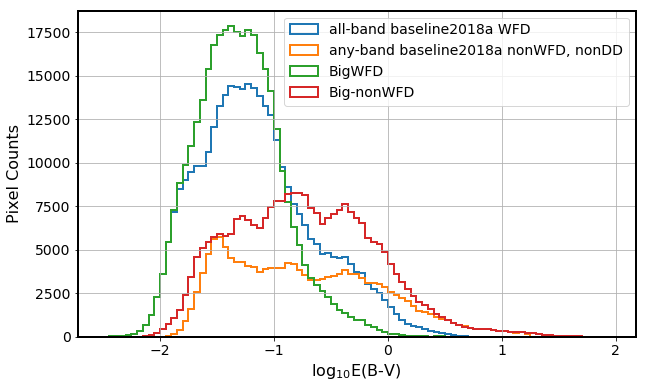}
  \caption{\humna{
    MW extinction distribution, E(B-V), for our proposed Big Sky WFD (green) and non-WFD (red) vs. \ttt{baseline2018a} WFD (blue) and non-WFD (orange). We see that the proposed WFD effectively excludes some regions of very high extinction. The extinction is derived from SFD maps \citep{sfd98} using \ttt{MAF} \citep{jones+14}.
    }
    }
    \label{fig:ebv}
\end{figure}

\noindent This solution has many benefits, including:\\
 $\bullet$ Avoidance of high extinction regions in WFD for extragalactic science; see \autoref{fig:ebv}.\\
$\bullet$ Larger area in WFD for the discovery of Milky Way structure and faint dwarf galaxies away from the Galactic Plane.\\
$\bullet$ Larger coverage of the Ecliptic Plane in WFD, increasing the discovery potential of Solar System objects.\\
$\bullet$ Inclusion of LMC and extended periphery in WFD, increasing the potential for discovery of its tidal debris and satellite system.\\
$\bullet$ Increased overlap with DESI and northern facilities for spectroscopic and photometric followup of LSST targets.

We note that this solution is proposed only for the 90\% time used for the main survey; 10\% time remains left over for Deep Drilling and mini-surveys.  By increasing the average number of visits in non-WFD areas it also increases the flexibility for mini-surveys.  
Greater coverage of e.g.\ the Galactic Plane could be accomplished by using a portion of the 10\% time not assigned to the main survey, or by trading visits from one part of the non-WFD sky to another.
As such, 
this white paper 
\eric{suggests} 
the basis for \phil{an initial} compromise solution to several of the concerns with the baseline cadence. \phil{Further iterations could be made to refine this solution, 
using the simple, fast simulation tool that we have provided.}

\clearpage
\vspace*{2em}

\clearpage
\section{Technical Description}

\subsection{High-level description\label{sec:high-level}}
We propose that the LSST survey consider a \textit{Big Sky} footprint that covers the full range of declinations visible from Cerro Pach\'{o}n, $-90\degree<\rm{Dec}<+32\degree$, which covers $\sim$ 30,000 $\deg^2$ in total.  We further propose that WFD be considered as the 18,000 $\deg^2$ with $|b|>15\degree$ (where $b$ is the Galactic latitude) and $-72.25\degree<\rm{Dec}<+12.4\degree$.  With no snaps and 30s exposure per visit, we adopt the constraint from \ttt{OpSim} that a 10-year survey will allow for 2.6M visits in total.  We assume that 90\% of these visits (2.34M) are available for WFD and mini-surveys (not including the DDF mini-survey).  We assign 825 visits to each WFD field, meeting the design goal defined in the LSST SRD trivially, and the remaining visits to the non-WFD fields in our Big Sky area. In the most basic plan, we distribute the remaining visits evenly over the non-WFD parts of the sky, yielding $\sim$ 250 visits per each non-WFD field.  The distribution of visits in the mini-surveys could further be optimized by fine-tuning the distribution of visits among non-WFD fields to cater to the science goals; there also remains, of course, 10\% time left to be used for Deep Drilling and mini-surveys. Note that this proposal does not address the {\it cadence} of either the WFD or the non-WFD areas, \phil{only the sky coverage and \humna{the} eventual 10-year depth.}


\vspace{.3in}

\subsection{Footprint -- pointings, regions and/or constraints}

Our requested footprint is depicted in \autoref{fig:coverage} \humna{(left panel)}, showing the Big Sky area that is covered by 4066 LSST fields. 2290 fields are assigned to WFD
, reconfiguring the WFD footprint away from the Galactic Plane, where stellar extinction and crowding is high. 

1776 fields 
are assigned to what we call the `extended footprint' mini-survey, with 254 visits per field over a 10-year period. These 1776 fields are concentrated in the Galactic Plane, Southern Celestial Pole, and northward of $\rm{Dec} = +12.4\degree$. In our proposed scenario, LSST would observe fields as far north as $\rm{Dec} = +32\degree$, the northernmost limit of the telescope. The number of filters and distribution of visits for the fields within the extended footprint is flexible and could be tailored to the science goals driving those extensions (e.g., Northern Ecliptic Spur, overlap with Euclid and DESI, microlensing).

\subsection{Image quality}
There are no special requirements for image quality.

\subsection{Individual image depth and/or sky brightness}
Ideally, the individual image depth would not vary significantly with declination beyond what is unavoidable due to seeing/airmass changes. \phil{With this uniform approach,} discovery and detection of moving objects with the same completeness level \phil{would be enabled} at any point within the WFD footprint and the extended footprint. 
Visits should be scheduled for optimal conditions per filter, as is currently done in the standard survey strategy.

\subsection{Co-added image depth and/or total number of visits}
There are no special requirements for co-added image depth. 
\subsection{Number of visits within a night}
To satisfy core Solar System science goals to identify moving objects in the WFD and 
NES region\humna{s}, we request (at least) two visits per night, separated by $\sim$~15-90 minutes. This requirement is driven by the LSST Moving Object Pipeline System (MOPS) requirements, defined in LSE-30\footnote{\url{http://ls.st/LSE-30}} and  LDM-156\footnote{\url{http://ls.st/LDM-156}}.

\subsection{Distribution of visits over time}
The distribution of visits over time should permit linking of moving objects via MOPS, but there are no other strong specifications for the distribution of visits over time. For MOPS to be efficient at detecting moving Solar System objects, three tracklets (a pair of images  in the same night, acquired no more than 90 minutes apart) acquired within 15 days are needed \citep[see LSE-30,  LDM-156, and][for further details]{2018Icar..303..181J}. 
A highlight of our proposed solution is that it will allow catering the distribution of visits in WFD vs. non-WFD regions to the guiding science goals\phil{; we refer the reader to the many other cadence white papers for details}.

\subsection{Filter choice}
The portion of the non-WFD footprint corresponding to the area currently covered by the NES mini-survey does not need  $u$, $y$, or $i$, band coverage as described in the NES Solar System Mini survey white paper 
\humna{(Schwamb et al.).} Potential trades on filter selection for the non-WFD footprint NES region to accommodate tensions with other requested cadences are described in Schwamb et al. All other areas of the WFD-footprint and the extended footprint would receive coverage in all LSST filters. 

\subsection{Exposure constraints}

Our proposed scenario to accommodate a wide variety of key science cases with a modified WFD footprint and extended footprint will $only$ be successful with one snap per visit. With two snaps per visit for {\it all} exposures, the loss in observing efficiency is large enough to reduce the number of visits allocated to the non-WFD footprint to 152, which is less than currently allocated to any mini-survey. 

\subsection{Other constraints}
None.

\subsection{Estimated time requirement}
As discussed in \autoref{sec:high-level}, we are 
proposing \eric{a configuration for}
90\% of the total visits available after dropping snaps. Since our proposal reconfigures the WFD and various mini-surveys (not including the DDFs), we are actually reconfiguring $\sim$ 95\% of the 
\eric{baseline} 
survey time. 
\eric{One} difference is that the WFD area is capped at 825 visits per field, while the equivalent current simulation \humna{(\ttt{kraken\_2014})} 
without snaps averages 999 visits per field in the WFD. More details on the approximate total time requested 
can be found using the guidelines in \href{https://github.com/lsst-pst/survey_strategy_wp}{our \ttt{GitHub} repo}.

In general, we have not detailed the impact of timing constraints beyond requiring the number of visits to be sufficient for MOPS; the impact on the time required can better addressed in a full \ttt{OpSim} run.  An overview of MOPS requirements can be found in \humna{LSE-30,  LDM-156, and \citet{2018Icar..303..181J}.}


\begin{table}[ht]
    \centering
    \begin{tabular}{l|l|l|l}
        \toprule
        Properties & Importance \hspace{.3in} \\
        \midrule
        Image quality                         &  2  \\  
        Sky brightness                        &  3  \\
        Individual image depth                &  1  \\
        Co-added image depth                  &  1  \\
        Number of exposures in a visit        &  1 \\  
        Number of visits (in a night)         &  2  \\ 
        Total number of visits                &  1 \\  
        Time between visits (in a night)      &  1  \\
        Time between visits (between nights)  &   1 \\
        Long-term gaps between visits         &   2  \\
        Other (please add other constraints as needed) & \\
        \bottomrule
    \end{tabular}
     \caption{{\bf Constraint Rankings:} Summary of the relative importance of various survey strategy constraints, ranked from 1=very important, 2=somewhat important, 3=not important. \phil{In our proposed strategy, the only important consideration is the total number of visits, which we seek to increase so as to serve as many science communities as possible; the number of exposures in a visit is hence also important, as we describe in the text.}}
        \label{tab:obs_constraints}
\end{table}

\subsection{Technical trades}
\begin{enumerate}
    \item We have evaluated trade-offs between survey footprint area and the number of visits per field as part of developing our tool, and have settled on a simple combination of an optimized footprint and the number of visits that meets various requirements. Please see more in \autoref{sec:motivation} and \autoref{sec:high-level}.
    \item The minimal frequency of observations should be such to enable minor planet discovery with 95$\%$ confidence (see LSE-30 and  LDM-156).
    \item Increasing the exposure time (and reducing the number of visits) would impact the number of visits available and the number allocated to the WFD. Given that the (median) number of WFD visits is set at the design goal, our suggested survey strategy is sensitive to these trades but has some wiggle room.
    \item Real-time exposure time optimization could be beneficial for maintaining individual image depth across the footprint as long as it does not significantly impact the total number of observations.
    \item Our proposal assigns nearly the minimum number of visits for both regions of the footprint (WFD and non-WFD). Reducing the number of visits significantly is likely to make this `compromise' solution less optimal. For example, the NES white paper (Schwamb et al.) requests about 255 visits per field, an optimized number that allows for the discovery, linkage, and monitoring of moving objects ranging from asteroids  to the  distant Outer Solar System bodies. Reducing the number of visits available in the non-WFD footprint below 255 will be problematic for the NES. 
\end{enumerate}

\clearpage
\section{Performance Evaluation}

\begin{figure}[!h]
    \centering
    \includegraphics[width=1\columnwidth, trim={0cm 0cm 0cm 0cm}, clip]{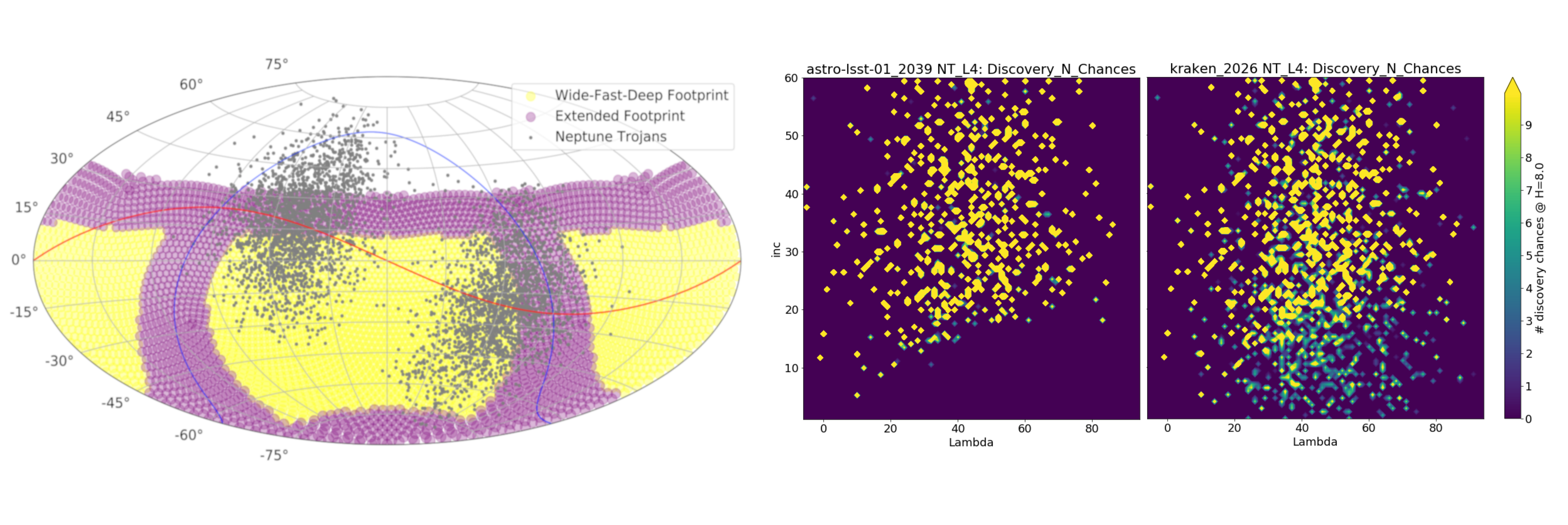}
    \caption{\textit{Left:} The on-sky positions of Neptune Trojans in 2022. \textit{Right:} The metric 
    \ttt{Discovery\_N\_Chances} 
    (\humna{which measures the} number of discovery opportunities of a particular object) as a function of orbital elements ``inclination'' (inc) and ``mean longitudes'' (Lambda). The \ttt{Opsim} run \humna{\ttt{astro-lsst-01\_2039}} 
    simulated the survey without Dec $> 0^\circ$ pointings, and \humna{\ttt{kraken\_2026}} 
    is the analog of our proposed survey with Dec $> 0^\circ$ coverage.}
    \label{fig:solarsystem}
\end{figure}

For the specifications of our proposal, the main metric is:
\begin{itemize}
    \item The total number of visits per field in the WFD and non-WFD areas.
\end{itemize}
Our proposed footprint can further be optimized by considering the following metrics:
\begin{itemize}
    \item 
    \eric{To evaluate the performance of our proposed strategy for dark energy science, we refer to the DESC WFD white paper \humna{(Lochner at al.) where the }
    authors identify a number of proxy metrics that are expected to correlate with their ultimate goal of maximizing the Dark Energy Task Force figure of merit (FoM).  These include 1) a Static Science Statistical FoM, which is emulated based on a WFD survey’s area and depth, 2) the average number of $i$-band visits, which reduces weak lensing systematics from PSF modeling, 3) the number of well-measured Type Ia supernovae obtained by excluding those at redshifts too high to offer acceptable data quality, 4) the number of strongly lensed Type Ia supernovae, and 5) the number of kilonovae detected.  }

    \item Metrics in the NES white paper (Schwamb et al.) are described in detail for evaluating the performance of WFD + Big Sky extension in the NES. Solar System Object Differential Completeness Goals for evaluating the performance of our proposed WFD footprint are detailed in the Community Observing Strategy Evaluation Paper. \citep[COSEP;][]{2017arXiv170804058L}. Figure~\ref{fig:solarsystem} shows an example of how this proposed footprint improves the discovery chance of Solar System objects. A special minor planet population called Neptune Trojans -- the asteroids \humna{that} co-orbit with  Neptune around its L4 and L5 Lagrangian points -- \humna{has} 
    wide and non-uniform on-sky distribution (Figure~\ref{fig:solarsystem}, left panel). Our proposed footprint greatly increase the discovery chances of the inclination $< 20^\circ$ L4 Trojans from 0 to several (Figure~\ref{fig:solarsystem}, right panel).
    \item Metrics in the Magellanic Clouds white paper (Olsen et al.) are described for evaluating the performance of Big Sky for mapping the structure of the Magellanic Clouds and their periphery, for identifying their dwarf satellite system, and for surveying their variable star and transient populations.  They fundamentally depend on the number of visits spent on the area south of Dec $\sim-60$.
    \item \humna{WFD and non-WFD area overlap with northern surveys. Since northern hemisphere facilities are generally found at latitudes between +20\degree and +35\degree, we can quantify the area overlap with these facilities by calculating the area overlap for Dec $>$ -20\degree. We find that our proposed WFD footprint will then see an overlap of 9277 deg$^2$ vs. \ttt{baseline2018a} overlap of 7789 deg$^2$, while our Big Sky non-WFD will see an overlap of 9078 deg$^2$ vs. \ttt{baseline2018a} overlap of 4870 deg$^2$. These numbers change to 8396 deg$^2$ vs. 6265 deg$^2$ for WFD and 5772 deg$^2$ vs 2744 deg$^2$ for non-WFD, if we only consider overlap in the regions with MW extinction $<$ 0.2 which is necessary for extragalactic science.
    }
    \item For the evaluation of Big Sky for MW halo streams and dwarf companions, a metric is the number of such objects that can be discovered.  The number is proportional to area, and depends on depth through the stellar luminosity function for the streams and dwarfs, the foreground contamination, the background galaxy contamination, and the integrated luminosity function of the streams and dwarfs.  One could also account for the ability to use proper motions to remove contamination.  While a quantitative metric needs further work, the comparison of the roughly equal number of dwarf satellites by SDSS (14,000 $\deg^2$) and DES (5,000 $\deg^2$) gives us an empirical measure with which to calibrate the metric.  \cite{Newton2018} find that LSST WFD should detect $\sim$100 Milky Way dwarfs.  Given the difficulty of finding dwarfs behind the Milky Way plane, this number would likely be reduced by $\sim25\%$ in the \ttt{baseline2018a} footprint compared to Big Sky.
    \item For the scientific value of the overlap of LSST with DESI, metrics are described in detail in Bolton et al.\ (``Maximizing the Joint Science Return of LSST and DESI").
\end{itemize}

\section{Special Data Processing}

The standard LSST data management pipelines will be able to process the observations taken in our proposed WFD and extended Big Sky footprint. The majority of science cases driving our proposed changes will use the LSST data management pipelines as described in the LSST science book \citep{2009arXiv0912.0201L}. We will essentially \eric{mirror special data processing requests submitted by other} 
white papers, including the NES white paper (Schwamb et al.) and the DESC WFD \humna{cadence} white paper (Lochner et al.).

\section{Acknowledgements}
 The authors thank and acknowledge the organizers of the  LSST Cadence Hackathon at the Flatiron Institute where the concept for this white paper was created. We also thank LSST Corporation and the Simons Foundation for their support.   
 
\bibliographystyle{aasjournal}
\bibliography{peace} 

\end{document}